\documentstyle[12pt]{article}

\skewchar\fivmi='177
\skewchar\sixmi='177
\skewchar\sevmi='177
\skewchar\egtmi='177
\skewchar\ninmi='177
\skewchar\tenmi='177
\skewchar\elvmi='177
\skewchar\twlmi='177
\skewchar\frtnmi='177
\skewchar\svtnmi='177
\skewchar\twtymi='177
\def\@magscale#1{ scaled \magstep #1}
\skewchar\fivsy='60
\skewchar\sixsy='60
\skewchar\sevsy='60
\skewchar\egtsy='60
\skewchar\ninsy='60
\skewchar\tensy='60
\skewchar\elvsy='60
\skewchar\twlsy='60
\skewchar\frtnsy='60
\skewchar\svtnsy='60
\skewchar\twtysy='60

\catcode`@=11
\def\un#1{\relax\ifmmode\@@underline#1\else
        $\@@underline{\hbox{#1}}$\relax\fi}
\catcode`@=12

\def\a{\alpha}
\def\b{\beta}

\def\d{\delta}
\def\e{\epsilon}

\def\l{\lambda}

\def\s{\sigma}

\def\dslash{\not{\hbox{\kern-2pt $\partial$}}}
\def\Dslash{\not{\hbox{\kern-4pt $D$}}}
\def\pslash{\not{\hbox{\kern-2.3pt $p$}}}
 \newtoks\slashfraction
 \slashfraction={.13}
 \def\slash#1{\setbox0\hbox{$ #1 $}
 \setbox0\hbox to \the\slashfraction\wd0{\hss \box0}/\box0 }

\font\ro=cmsy10                          
\def\kcr{{\hbox{\ro \char'170}}}                
\def\ktl{{\hbox{\ro \char'170}}}        
\def\ktr{{\hbox{\ro \char'170}}}        
\def\kbl{{\hbox{\ro \char'170}}}        
\def\kbr{{\hbox{\ro \char'170}}}        

\def\plpl{\raise-2pt\hbox{$\raise3pt\hbox{$_+$}\hskip-6.67pt\raise0.0pt
\hbox{$^+$}\hskip 0.01pt$}}
\def\mimi{\raise-2pt\hbox{$\raise3pt\hbox{$_-$}\hskip-6.67pt\raise0.0pt
\hbox{$^-$}\hskip 0.01pt$}}

\def\bo{{\raise.15ex\hbox{\large$\Box$}}}               
\def\pa{\partial}                                       
\def\TH{{\raise.2ex\hbox{$\displaystyle \bigodot$}\mskip-4.7mu \llap H \;}}
\def\face{{\raise.2ex\hbox{$\displaystyle \bigodot$}\mskip-2.2mu \llap {$\ddot
        \smile$}}}                                      

\def\sp#1{{}^{#1}}                              
\def\leftrightarrowfill{$\mathsurround=0pt \mathord\leftarrow \mkern-6mu
        \cleaders\hbox{$\mkern-2mu \mathord- \mkern-2mu$}\hfill
        \mkern-6mu \mathord\rightarrow$}
\def\dvec#1{\vbox{\ialign{##\crcr
        \leftrightarrowfill\crcr\noalign{\kern-1pt\nointerlineskip}
        $\hfil\displaystyle{#1}\hfil$\crcr}}}           

\def\fracm#1#2{\hbox{\large{${\frac{{#1}}{{#2}}}$}}}
\def\frac#1#2{{\textstyle{#1\over\vphantom2\smash{\raise.20ex
        \hbox{$\scriptstyle{#2}$}}}}}                   
\def\sfrac#1#2{{\vphantom1\smash{\lower.5ex\hbox{\small$#1$}}\over
        \vphantom1\smash{\raise.4ex\hbox{\small$#2$}}}} 
\def\bfrac#1#2{{\vphantom1\smash{\lower.5ex\hbox{$#1$}}\over
        \vphantom1\smash{\raise.3ex\hbox{$#2$}}}}       
\def\afrac#1#2{{\vphantom1\smash{\lower.5ex\hbox{$#1$}}\over#2}}    

\newskip\humongous \humongous=0pt plus 1000pt minus 1000pt
\def\caja{\mathsurround=0pt}
\def\eqalign#1{\,\vcenter{\openup2\jot \caja
        \ialign{\strut \hfil$\displaystyle{##}$&$
        \displaystyle{{}##}$\hfil\crcr#1\crcr}}\,}
\newif\ifdtup

\def\ref#1{$\sp{#1)}$}

\parskip=\medskipamount                 
\thicklines                         

\def\oldheadpic{                                
        \setlength{\unitlength}{.4mm}
        \thinlines
        \par
        \begin{picture}(349,16)
        \put(325,16){\line(1,0){4}}
        \put(330,16){\line(1,0){4}}
        \put(340,16){\line(1,0){4}}
        \put(335,0){\line(1,0){4}}
        \put(340,0){\line(1,0){4}}
        \put(345,0){\line(1,0){4}}
        \put(329,0){\line(0,1){16}}
        \put(330,0){\line(0,1){16}}
        \put(339,0){\line(0,1){16}}
        \put(340,0){\line(0,1){16}}
        \put(344,0){\line(0,1){16}}
        \put(345,0){\line(0,1){16}}
        \put(329,16){\oval(8,32)[bl]}
        \put(330,16){\oval(8,32)[br]}
        \put(339,0){\oval(8,32)[tl]}
        \put(345,0){\oval(8,32)[tr]}
        \end{picture}
        \par
        \thicklines
        \vskip.2in}
\def\oldtitle#1#2#3#4{\oldheadpic\begin{center}\vglue.5in{\large\bf #1}\\[.6in]
        {#2}\\[.1in] {\it Department of Physics and Astronomy}\\
        {\it University of Maryland, College Park, MD 20742}\\[.6in]
        Physics Publication \#{#3}\\ {#4}\\[1.5in] {\bf ABSTRACT}\\[.1in]
        \end{center} \begin{quotation}}                 
\def\oldTitle#1#2#3#4#5#6#7{\oldheadpic\begin{center} \vglue .4in
        {\large\bf #1}\\[.4in]
        {#2}\\[.1in] {\it Department of Physics and Astronomy}\\
        {\it University of Maryland, College Park, MD 20742}\\[.1in]
        {#3}\\[.1in] {\it {#4}}\\ {\it {#5}}\\[.4in]
        Physics Publication \#{#6}\\ {#7}\\[.5in] {\bf ABSTRACT}\\[.1in]
        \end{center} \begin{quotation}}                 
\def\border{                                            
        \setlength{\unitlength}{1mm}
        \newcount\xco
        \newcount\yco
        \xco=-21
        \yco=12
        \begin{picture}(140,0)
        \put(\xco,\yco){$\ktl$}
        \advance\yco by-1
        {\loop
        \put(\xco,\yco){$\kcr$}
        \advance\yco by-2
        \ifnum\yco>-240
        \repeat
        \put(\xco,\yco){$\kbl$}}
        \xco=158
        \yco=12
        \put(\xco,\yco){$\ktr$}
        \advance\yco by-1
        {\loop
        \put(\xco,\yco){$\kcr$}
        \advance\yco by-2
        \ifnum\yco>-240
        \repeat
        \put(\xco,\yco){$\kbr$}}
        \put(-20,13){\tiny University of Maryland Elementary Particle
Physics University of Maryland Elementary Particle Physics University of
Maryland Elementary Particle Physics}
        \put(-20,-241.5){\tiny University of Maryland Elementary
Particle Physics University of Maryland Elementary Particle Physics
University of Maryland Elementary Particle Physics}
        \end{picture}
        \par\vskip-8mm}
\def\bordero{                                           
        \setlength{\unitlength}{1mm}
        \newcount\xco
        \newcount\yco
        \xco=-31
        \yco=12
        \begin{picture}(140,0)
        \put(\xco,\yco){$\ktl$}
        \advance\yco by-1
        {\loop
        \put(\xco,\yco){$\kclr}
        \advance\yco by-2
        \ifnum\yco>-240
        \repeat
        \put(\xco,\yco){$\kbl$}}
        \xco=151
        \yco=12
        \put(\xco,\yco){$\ktr$}
        \advance\yco by-1
        {\loop
        \put(\xco,\yco){$\kcr$}
        \advance\yco by-2
        \ifnum\yco>-240
        \repeat
        \put(\xco,\yco){$\kbr$}}
        \put(-20,12){\ooo
bacdefghidfghghdhededbihdgdfdfhhdheidhdhebaaahjhhdahba

hgdedge
   hgfdiehhgdigicba}
        \put(-20,-241.5){\ooo
ababaighefdbfghgeahgdfgafagihdidihiidhiagfedhadbfd

ecdcdfa
   gdcbhaddhbgfchbgfdacfediacbabab}
        \end{picture}
        \par\vskip-8mm}
\def\headpic{                                           
        \indent
        \setlength{\unitlength}{.4mm}
        \thinlines
        \par
        \begin{picture}(29,16)
        \put(165,16){\line(1,0){4}}
        \put(170,16){\line(1,0){4}}
        \put(180,16){\line(1,0){4}}
        \put(175,0){\line(1,0){4}}
        \put(180,0){\line(1,0){4}}
        \put(185,0){\line(1,0){4}}
        \put(169,0){\line(0,1){16}}
        \put(170,0){\line(0,1){16}}
        \put(179,0){\line(0,1){16}}
        \put(180,0){\line(0,1){16}}
        \put(184,0){\line(0,1){16}}
        \put(185,0){\line(0,1){16}}
        \put(169,16){\oval(8,32)[bl]}
        \put(170,16){\oval(8,32)[br]}
        \put(179,0){\oval(8,32)[tl]}
        \put(185,0){\oval(8,32)[tr]}
        \end{picture}
        \par\vskip-6.5mm
        \thicklines}
\def\title#1#2#3#4{\border\headpic {\hbox to\hsize{#4 \hfill UMDEPP #3}}\par
        \begin{center} \vglue .5in {\large\bf #1}\\[.6in]
        {#2}\\[.1in] {\it Department of Physics and Astronomy}\\
        {\it University of Maryland, College Park, MD 20742}\\[1.5in]
        {\bf ABSTRACT}\\[.1in] \end{center} \begin{quotation}}  
\def\Title#1#2#3#4#5#6#7{\border\headpic
        {\hbox to\hsize{#7 \hfill UMDEPP #6}}\par
        \begin{center} \vglue .4in {\large\bf #1}\\[.4in]
        {#2}\\[.1in] {\it Department of Physics and Astronomy}\\
        {\it University of Maryland, College Park, MD 20742}\\[.1in]
        {#3}\\[.1in] {\it {#4}}\\ {\it {#5}}\\[.5in] {\bf ABSTRACT}\\[.1in]
        \end{center} \begin{quotation}}                 
\def\endtitle{\end{quotation}\newpage}                  

\def\sect#1{\bigskip\medskip \goodbreak \noindent{\bf {#1}} \nobreak \medskip}

\begin{document}

\def\fracm{\frac}
\def\dvm{\raisebox{-.45ex}{\rlap{$=$}} }
\def\DM{{\scriptsize{\dvm}}~~}

\def\lin{\vrule width0.5pt height5pt depth1pt}
\def\dpx{{{ =\hskip-3.75pt{\lin}}\hskip3.75pt }}

\border\headpic {\hbox to\hsize{October 1994 \hfill UMDEPP 95-060}}\par
\begin{center}
\vglue .4in
{\large\bf  MANIFEST (4,0) SUPERSYMMETRY, SIGMA MODELS \\
AND THE \\
ADHM INSTANTON CONSTRUCTION \footnote{Research supported
by NSF grant \# PHY-93-41926}
${}^,$ \footnote {Supported in part by NATO Grant CRG-93-0789}
  }\\[.2in]
S. James Gates, Jr.\footnote{gates@umdhep.umd.edu} and
Lubna Rana\footnote{lubna@umdhep.umd.edu} \\[.1in]
{\it Department of Physics\\
University of Maryland at College Park\\
College Park, MD 20742-4111, USA}
\\[3.0in]

{\bf ABSTRACT}\\[.1in]
\end{center}
\begin{quotation}

Utilizing (4,0) superfields, we discuss aspects of supersymmetric
sigma-models and the ADHM construction of instantons \' a la
Witten.
\endtitle

\sect{I. Introduction}

Recently \cite{WTN} an argument has been given that suggests that
(4,0) sigma models provide a natural setting in which to discuss
the instanton construction of Atiyah, Drinfeld, Hitchin and Manin
\cite{ADHM}.   As shown in reference \cite{WTN}, there is an elegant
relation between (4,0) supersymmetric theories of scalar and spinor
multiplets and the prior work on the construction of instantons.  In
his presentation, Witten, pointed out the need to study the extent to
which it is possible to generalize these new results.  This is the main
purpose of this presentation.  Namely, it is our goal to write
out the most general possible model along the following lines.
Foremost, the (4,0) action must involve scalar and spinor multiplets.
Secondly, the spinor multiplets must have (generalized
Yukawa-type) interactions.  Thirdly, we demand the presence of
manifest (4,0) supersymmetry at all stages of investigation.

Coincidently, just prior to the appearance of the work in \cite{WTN} , we
had exactly developed the requisite tools \cite{GRana} for this study
while investigating aspects of N-extended supersymmetry within the confines
of 1D supersymmetric quantum mechanical models. In fact, for a fixed
N, there is a one-to-one correspondence between 1D models and 2D
``heterotic'' models. This should come as no surprise since on-shell
``heterotic'' models are just 1D models.  In fact, as we shall shortly
see, (4,0) unidexterous supersymmetry has an unimagined richness in
the number of representations from which to build models along the
lines described above.  Furthermore all of the formulations that
we shall use possess complete off-shell representations containing all
the necessary auxiliary fields.  This latter point is particularly
important as it permits us to easily establish the
explicit forms of the interactions between the scalar and spinor
multiplets.

\sect{II. Free (4,0) Scalar Multiplets}

One unexpected result found during the investigation of 1D, N = 4
supersymmetric models \cite{GRana} was the surprisingly large
number of such representations.  There are, to our knowledge, four
(4,0) scalar multiplets.  This is an example of the phenomenon of
variant superfield representations \cite{SG1}.

The first scalar multiplet contains spin-0 fields, ${\cal A}$, ${\cal
B}$ and spinor fields $\psi^- {}^{j}$.  The supersymmetry variations
of the component
$$ \eqalign{ \d_{Q} {\cal A}~~~ =~~~ & 2 {\e}^+ {}^i C_{ij} \psi^-
             {}^{j} ~~~, \cr
             \d_{Q} {\cal B}~~~ =~~~ & i 2 {\bar \e}^+ {}_{i} \psi^-
             {}^{i} ~~~,\cr
             \d_{Q} \psi^- {}^{i}~~~ =~~~ & i{\bar \e}^+ {}_{j} C^{ij}
             \pa_{\dpx} {\cal A} ~-~ {\e}^+ {}^{i} \pa_{\dpx} {\cal B}
             ~~~. }
\eqno(2.1)$$
We refer to this as the (4,0) SM-I (scalar multiplet one) theory. The
free kinetic energy for the fields (written below) form a supersymmetric
invariant given by
$$ {\cal S}_{{\rm SM-I}} = \int d^2 \s \left[ ~ ( \pa_{\dpx} {\cal A} )
\left( \pa_{\DM} {\bar{\cal A}} \right) + ( \pa_{\dpx} {\cal B} )
\left( \pa_{\DM} {\bar{\cal B}} \right) + i 2 {\bar \psi }^{-} {}_{i}
\left( \pa_{\DM} \psi^- {}^{i}  \right) ~ \right]   ~~~. \eqno(2.2)$$

The second scalar multiplet has the component structure ($  \phi,~
\phi_{i}{}^{j}  ,~  \lambda^- {}_{i} $) where the first two fields are
bosons and the latter a fermion. The supersymmetry variations are
$$ \eqalign{ \d_{Q} \phi~~~ =~~~& i  \e^+ {}^i \lambda^- {}_i ~+~ i
{\bar \e}_-  {}_{i} {\bar{\lambda}^-} {}^{i} ~~~, \cr
\d_{Q} \phi_{i}{}^{j}~~~ =~~~& 2 \left( \e^+ {}^{j} \lambda^- {}_{i} -
\frac12 \d_{i} {}^{j} \e^+ {}^{k} \lambda^- {}_{k} \right) - 2 \left(
{\bar \e}^+ {}_{i} {\bar{\lambda}^-}{}^{j} - \frac12 \d_{i}{}^{j}
{\bar \e}^+ {}_{k} {\bar{\lambda}^-}{}^{k} \right) ~~~, \cr
\d_{Q} \lambda^- {}_{i}~~~=~~~&- {\bar \e}^+ {}_{i} \pa_{\dpx}
\phi ~-~ i  {\bar \e}^+ {}_k \pa_{\dpx} \phi_{i}{}^{k} ~~~. } \eqno(2.3) $$
We refer to this as the (4,0) SM-II (scalar multiplet two) theory and
the following action is invariant under these variations
$$ {\cal S}_{{\rm SM-II}} = \int d^2 \s \left[ ~  (
\pa_{\dpx} \phi ) \left( \pa_{\DM} \phi \right) + \fracm 12 ( \pa_{\dpx}
\phi_{i}{}^{j} )  \left( \pa_{\DM} \phi_{j}{}^i  \right) + i
2 ( {\bar \lambda}^- {}^{i} \pa_{\DM} \lambda^- {}_{i} )  ~ \right] ~~~.
\eqno(2.4) $$

The third scalar multiplet has the component structure ($  {\cal A}_i ,~
\rho^-  ,~  \pi^-  $) where the first field is a boson and the latter two
are fermions. The supersymmetry variations are
$$ \eqalign{
\d_{Q} {\cal A}_i ~~~ =~~~ &  C_{ij} \, \e^+ {}^{j} \, \pi^-  ~+~
{\bar \e}^+ {}_i \rho^-  ~~~, \cr
\d_{Q} \rho^- ~~~ =~~~ & - i 2 \, \e^+ {}^{i} \pa_{\dpx} {\cal A}_i ~~~, \cr
\d_{Q} \pi^- ~~~ =~~~ & i2 \, C^{ij} \, {\bar \e}^+ {}_{i} \, \pa_{\dpx}
{\cal A}_j  ~~~. }
\eqno(2.5)$$

We refer to this as the (4,0) SM-III (scalar multiplet three) theory and
the following action is invariant under these variations
$$ {\cal S}_{{\rm SM-III}} = \int d^2 \s \left[ ~ ( \pa_{\dpx} {\bar {\cal
A}}^i )  \left( \pa_{\DM}{\cal A}_i \right) + i \fracm12 {\bar{\rho}}^-
\pa_{\DM} {\rho}^- + i \fracm12 {\bar{\pi}}^- \pa_{\DM} {\pi}^- ~ \right] ~~~.
\eqno(2.6) $$

The final scalar multiplet (known to us) has the component structure ($
{\cal B}_i ,~ \psi^-  ,~ \psi^- {}_i {}^{j}  $) where the first field is
a boson and the latter two are fermions. The supersymmetry variations are
$$ \eqalign{
\d_{Q} {\cal B}_{i}~~~ =~~~ & {\bar \e}^+ {}_i \, \psi^- ~-~ i 2 \,
{\bar \e}^+ {}_{j} \, \psi^- {}_i {}^{j} ~~~, \cr
\d_{Q} \psi^-~~~ =~~~ & - i \e^+ {}^{i} \, \pa_{\dpx} {\cal B}_{i} +
 i {\bar \e}^+ {}_i \pa_{\dpx} {\bar{\cal B}}^{i}  ~~~, \cr
\d_{Q} \psi^- {}_i {}^{j}~~~ =~~~ & \left( \e^+ {}^{j} \, \pa_{\dpx}
{\cal B}_{i} - \frac12 \d_i {}^{j} \, \e^+ {}^{k} \, \pa_{\dpx} {\cal
 B}_{k} \right) + \left( {\bar \e}^+ {}_i \, \pa_{\dpx} {\bar{\cal B}}^{j}
- \frac12 \d_i {}^{j} {\bar \e}^+ {}_k \, \pa_{\dpx} {\bar{\cal B}}^{k}
\right)   ~~~. } \eqno(2.7) $$

We refer to this as the (4,0) SM-IV (scalar multiplet four) theory and
the following action is invariant under these variations
$$ {\cal S}_{{\rm SM-IV}} = \int d^2 \s \left[ ~ ( \pa_{\dpx} {\bar {\cal
B}}^i )  \left( \pa_{\DM}{\cal B}_i \right) +  i \fracm12 {\psi}^-
\pa_{\DM} {\psi}^- +  i  {\psi}^- {}_{i} {}^j \pa_{\DM} {\psi}^- {}_{j} {}^i
{}~ \right] ~~~.
\eqno(2.8) $$

We summarize these results in the table below.
\begin{center}
\renewcommand\arraystretch{1.2}
\begin{tabular}{|c|c|c| }\hline
${\rm Multiplet}$  & ${\rm Spin}$-${\rm 0~SU(2)~Rep}$ & ${\rm Spin}$-${\fracm
12}
{\rm~SU(2)~Rep}$  \\ \hline  \hline
${\rm SM-I}$ & ${~~~~4s~~~~}$ &  ${~~\fracm 12 ~~~}$ \\ \hline
${\rm SM-II}$ & ${~~1s1p~~~~}$ &  ${~~\fracm 12 ~~~}$ \\ \hline
${\rm SM-III}$ & ${~~\fracm 12 ~~~}$ &   ${~~~~4s~~~~}$ \\ \hline
${\rm SM-IV}$ &  ${~~\fracm 12 ~~~}$ &  ${~~1s1p~~~~}$  \\ \hline
\end{tabular}
\end{center}
\vskip.2in
\centerline{{\bf Table I}}

\sect{III. Free (4,0) Spinor Multiplets}

Similarly, for ``minus spinor multiplets'' (MSM) the same large number
of theories make their appearance.  There are four such multiplets.
The fact that there are precisely four minus spinor multiplets is no
accident. Each of these multiplets can be paired with one of the scalar
multiplets by a recently recognized type \cite{GRana} of ``fermionic
duality.''
\begin{center}
{\bf Fermionic Duality Pairs}
\end{center}
\begin{center}
\renewcommand\arraystretch{1.2}
\begin{tabular}{|c|c| }\hline
${\rm Scalar~Multiplet}$  & ${\rm Fermionic~Dual~Multiplet}$
\\ \hline  \hline
 $~~{\rm SM-I}~~$ & ${\rm MSM-I} $  \\ \hline
$~~{\rm SM-II}~~ $ & ${\rm MSM-II} $  \\ \hline
$~~{\rm SM-III}~~ $ & ${\rm MSM-III}$  \\ \hline
$~~{\rm SM-IV}~~ $ & ${\rm MSM-IV} $  \\ \hline
\end{tabular}
\end{center}
\vskip.2in
\centerline{{\bf Table II}}

The component fields of our first minus spinor multiplet are ($
\rho^+ {}_i ,~ {\cal F}, ~ {\cal H}$) and their supersymmetry
variations are just;
$$ \eqalign{
\d_{Q} \rho^+ {}_i ~~=~~~& - C_{i j}  \e^+ {}^{j} {\cal F} ~-~ i
{\bar \e}^+ {}_{i}  {\cal H} ~~~, \cr
\d_{Q} {\cal F} ~~~=~~~& -i 2  C^{ij}{\bar \e}^+ {}_i \pa_{\dpx}
\rho^+ {}_j~~~, \cr
\d_{Q} {\cal H} ~~~=~~~& 2 \e^+ {}^{i} \pa_{\dpx} \rho^+ {}_i ~~~. \cr
}  \eqno(3.1) $$
We refer to this as the (4,0) MSM-I (minus spinor multiplet one) theory.
It has a supersymmetrically invariant action given by
$${\cal S}  = \int d^2 \s \left[ ~ i  {\bar \rho}^+  {}^{i}
(\pa_{\DM} \rho^+ {}_{i} )  + \frac{1}{2}  {\cal F} {\bar{{\cal F}}}  +
\frac{1}{2}  {\cal H} {\bar{{\cal H}}} ~ \right] ~~~.
\eqno(3.2)$$

The next minus spinor is composed of component fields ($ \chi^+ {}_i ,~
H ,~  H_{i}{}^{j}  $) together with the supersymmetry variations
given by,
$$ \eqalign{
\d_{Q} \chi^+ {}_i ~~~=~~~&-{\frac12} {\bar \e}^+ {}_i H ~-~ i 2 {\bar
\e}^+ {}_{k} H_{i}{}^{k} ~~~,  \cr
\d_{Q} H ~~~ =~~~& i 2 \e^+ {}^i \pa_{\dpx} \chi^+ {}_i ~+~ i 2
{\bar \e}^+ {}_i \pa_{\dpx} {\bar{\chi}^+} {}^{i}  ~~~, \cr
\d_{Q} H_{i}{}^{j}~~~ =~~~& \left( \e^+ {}^j \pa_{\dpx} \chi^+ {}_i - \frac12
\d_{i}{}^{j} \e^+ {}^{k} \pa_{\dpx} \chi^+ {}_k \right) - \left( {\bar \e}^+
{}_{i} \pa_{\dpx} {\bar{\chi}^+} {}^{j} - \frac12 \d_{i}{}^{j} {\bar \e}^+ {}_k
\pa_{\dpx} {\bar{\chi}^+} {}^{k} \right)  ~~~. \cr
}  \eqno(3.3) $$
We refer to this as the (4,0) MSM-II (minus spinor multiplet two) theory
and its invariant free action is,
$${\cal S}  = \int  d^2 \s \left[ ~ i  {\bar \chi}^+ {}^{i} (\pa_{\dpx}
\chi^+ {}_{i} )  + \fracm 18  H^2  +  H_{i}{}^{j}  H_{j}{}^{i}
{}~ \right] ~~~.
\eqno(3.4)$$

Component fields of the third multiplet consist of ($\a^+,~ \b^+ , ~
{\cal C}_i  $) which have the following supersymmetry variations
$$ \eqalign{
\d_{Q} \a^+ ~~~ =~~~ & - i 2 \, {\e}^+ {}^{i}  \, {\cal C}_i ~~~, \cr
\d_{Q} \b^+ ~~~ =~~~ & i2 \, C^{ij} \, {\bar \e}^+ {}_i \, \, {\cal C}_j
 ~~~,  \cr
\d_{Q} {\cal C}_i ~~~ =~~~ &  C_{ij} \, \e^+ {}^{j} \, \pa_{\dpx} \b^+ ~+~
{\bar \e}^+ {}_i \pa_{\dpx} \a^+  ~~~. \cr  } \eqno(3.5)$$
We refer to this as the (4,0) MSM-III (minus spinor multiplet three) theory
with invariant action given by,
$${\cal S}  = \int  d^2 \s \left[ ~ i  {\bar{\a}^+ } (\pa_{\dpx} \a^+ ) + i
{\bar{\b}^+ } (\pa_{\dpx} \b^+ )  - 2 {\bar {\cal C}}^{i} {{\cal C}}_{i}
{}~ \right] ~~~.
\eqno(3.6)$$

The final such multiplet has fields ($\chi^+ ,~  \chi^+ {}_i {}^{j} ,~
{\cal F}_{i} $) whose supersymmetry variations explicitly take the form
$$ \eqalign{ ~~~~~~~~~~
\d_{Q} \chi^+ ~~~ =~~~ & - i \e^+ {}^{i} \,  {\cal F}_{i} ~+~ i
{\bar \e}^+ {}_i  {\bar{\cal F}}^{i}  ~~~, \cr
\d_{Q} \chi^+ {}_i {}^{j}~~~ =~~~ & \left( \e^+ {}^j \,  {\cal F}_{i} - \frac12
\d_i {}^{j} \, \e^+ {}^k \,  {\cal F}_{k} \right)
+ \left( {\bar \e}^+ {}_i \,  {\bar{\cal F}}^{j}
- \frac12 \d_i {}^{j} {\bar \e}^+ {}_k \,  {\bar{\cal F}}^{k} \right) ~~~, \cr
\d_{Q} {\cal F}_{i}~~~ =~~~ & {\bar \e}^+ {}_i \, \pa_{\dpx} \chi^+ ~-~ i 2 \,
{\bar \e}^+ {}_j \, \pa_{\dpx} \chi^+ {}_i {}^{j} ~~~. \cr
}  \eqno(3.7) $$
We refer to this as the (4,0) MSM-IV (minus spinor multiplet four) theory.
The action left invariant under these supersymmetry variations is
$${\cal S}  = \int  d^2 \s \left[~  i \fracm 12 {{\chi}}^+ (\pa_{\dpx} \chi^+ )
+ i {{\chi}^+ }_{i}{}^{j} (\pa_{\dpx} \chi^+ {}_{j}{}^{i} )
  +  {\bar {\cal F}}^{i} {{\cal F}}_{i}
 ~\right] ~~~.
\eqno(3.8)$$

In closing, we note that our manifest (4,0) supersymmetric formulation shows
one interesting modification to the formulation of reference \cite{WTN}.
In his work, Witten introduced on-shell spinors (in his notation $\l^a_+$,
see his equation (2.8)) where the number of these fields was some arbitrary
integer $a = 1,...,n$.  Here we see that the number of real component
spinors must always be equal to a multiple of four. In fact, if $n$ is
not a multiple of four, the resulting theory is not even (4,0)
supersymmetric!

\sect{IV. Generalized ADHM (4,0) Mass \& Yukawa Interactions}

In the previous two sections, we have explicitly seen the great abundance
of (4,0) supersymmetric scalar and spinor multiplets. There are a total of
eight different multiplets that we must consider in the class of actions
of our interest.  The general member of this class has the form,
$$ {\cal S} ~=~ {\cal S}_{\rm Free} ~+~ {\cal S}_{\rm Mass}
{}~+~ {\cal S}_{\rm Yukawa} ~~~, \eqno(4.1) $$
where ${\cal S}_{\rm Free}$ is any linear combination of the free actions
that we have seen in the previous two sections.  Before proceeding with
our considerations, it is useful to note that the problem of introducing
the most general potential in (p,0) supersymmetric models has been
studied previously in terms of (1,0) superfields \cite{Hull}. In the
remainder of this section, we focus our attention on finding the most
general mass and interaction Lagrangian consistent with the proposal
of Witten.  It simplifies our discussion in that we need only consider
mass and ordinary Yukawa-type couplings.  A priori, 2D field theory
admits generalized Yukawa-type couplings of the form $B^n \times F
\times F$ where $B$ denotes a bosonic field while $F$ denotes a
fermionic one. In general $n$ can be an arbitrary integer.  However,
due to supersymmetry, the restriction that the scalar potential be of
no greater than degree (2,2), restricts us to the cases of $n = 0,1$.

We first consider the $n = 0$ mass terms. It is simple to see that
there is a unique SM-I mass term given by,
$$ {\cal S}_{\rm M_1} ~=~ {\rm M_1}  \int d^2 \s ~\left[ ~ \psi^- {}^i
\rho^+ {}_i ~+~ \fracm 12 {\cal A} {\cal F} ~-~ \fracm 12 {\cal B}
{\cal H} ~ \right] ~+~ {\rm h.} {\rm c.} ~~~, \eqno(4.2) $$
while the SM-II mass term is just
$$ {\cal S}_{\rm M_2} ~=~ {\rm M_2}  \int d^2 \s ~ \left[ ~ i \l^- {}^i
{\bar \chi}^+ {}_i ~+~i {\bar \l}^- {}^i \chi^+ {}_i ~-~ \fracm 12
\phi H  ~-~  \phi_{i}{}^{j} H_{j}{}^{i} ~ \right]  ~~~. \eqno(4.3)
$$
Continuing, we have the SM-III mass term
$$ {\cal S}_{\rm M_3} ~=~ {\rm M_3} \int d^2 \s ~\left[ ~  i \fracm 12
{\pi}^-   \a^+  ~-~ i \fracm 12 {\rho}^- \b^+ ~+~ C^{i j}  {\cal A}_i
{\cal C}_j ~ \right]  ~+~ {\rm h.} {\rm c.} ~~~, \eqno(4.4)
$$
and finally the SM-IV mass term given by
$$ {\cal S}_{\rm M_4} ~=~ {\rm M_4} \int d^2 \s ~\left[ ~
  i {{\psi}}^-  \chi^+  ~+~ i 2  {{\psi}^- }_{i}{}^{j}
\chi^+ {}_{j}{}^{i} ~-~  {\cal B}_i {\bar {\cal F}}^i
 ~-~  {\bar {\cal B}}^i   {\cal F}_i~ \right]  ~~~. \eqno(4.5)
$$
It is of interest to note that one linear combination of these
mass terms corresponds precisely to the N = 4 (i.e. (4,4)) mass
term that was recently discussed \cite{Q}.  This is in accord
with a conjecture of Witten \cite{WTN1} that in the limit of
vanishing instanton size there should correspond a full N = 4
ADHM sigma model. The ADHM sigma model mass term for which this
is true is the sum of ${\cal S}_{\rm M_1}$ and ${\cal S}_{\rm M_2}$.
The reason for this is obvious, the twisted-I multiplet in
reference \cite{Q} is the sum of SM-I and MSM-II and the twisted-II
multiplet in reference \cite{Q} is the sum of SM-II and MSM-I!

Our manifest (4,0) supersymmetric formulation of the multipets may be
added together without changing the underlying (4,0) supersymmetry algebra
which always takes the form (in terms of the D-algebra),
$$\eqalign{
\{ {D}_{+ i }~,~ {D}_{+ j } \} ~=~ 0 ~~~&,~~~
\{ {D}_{+ i }~,~ {\bar D}_+ {}^j \} ~=~ i 2 \d_i {}^j  \pa_{\dpx}  ~~~.
} \eqno(4.6)  $$
This algebra is realized on all of the component fields without
the use of any equations of motion.

The next step in our analysis will take advantage of the fact that
we have already found the mass terms. We re-write these as
$$ {\cal S}_{\rm Y_1} ~=~   \int d^2 \s ~\left[ ~ s^- {}^i
\rho^+ {}_i ~+~ \fracm 12 r_{\cal A} {\cal F} ~-~ \fracm 12
t_{\cal B} {\cal H} ~ \right] ~+~ {\rm h.} {\rm c.} ~~~,
\eqno(4.7) $$

$$ {\cal S}_{\rm Y_2} ~=~   \int d^2 \s ~ \left[ ~ i q^- {}_i
{\bar \chi}^+ {}^i ~+~i {\bar q}^- {}^i \chi^+ {}_i ~-~ \fracm 12
r_H  H  ~-~  t_{i}{}^{j} H_{j}{}^{i} ~ \right]  ~~~, \eqno(4.8)
$$

$$ {\cal S}_{\rm Y_3} ~=~  \int d^2 \s ~\left[ ~i \fracm 12
p^-   \a^+  ~-~ i \fracm 12 h^- \b^+  ~+~ L^i {\cal C}_j
 ~ \right] ~+~ {\rm h.} {\rm c.} ~~~,   \eqno(4.9) $$

$$ {\cal S}_{\rm Y_4} ~=~  \int d^2 \s ~\left[ ~
  i {K}^-  \chi^+  ~+~ i 2  {K^- }_{i}{}^{j}
\chi^+ {}_{j}{}^{i} ~-~ T_i {\bar {\cal F}}^i ~-~ {\bar T}^i
{\cal F}_i  ~ \right]  ~~~. \eqno(4.10)
$$
With an appropriate identification of the coefficients these reduce back
to the mass terms. However, we can use these expressions in a different
way to search for the $n = 1$ Yukawa terms!  Each of the sets of
functions that appear in (4.7 - 4.10) constitute a ``section'' along the
lines defined in reference \cite{Hull}.  For example, a very simple
choice of these sections corresponds to the introduction of the (4,0)
cosmological term.  Interestingly enough, if we maintain SU(2)
covariance, the cosmological term only exists for the sections in
(4.7) and (4.8). The cosmological term corresponds to
$$ \eqalign{
 ~~~~~~&( r_{\cal A},~ t_{\cal B},~ s^- {}^i ) ~\equiv~
( c_1,~ c_2 ,~ 0 ) ~~~, \cr
&( r_H ,~ t_{i}{}^{j},~  q^- {}_i ) ~\equiv~ ( c_3,~ 0 ,~ 0 )
{}~~~. } \eqno(4.10)
$$
for arbitrary complex constants $c_1$ , $c_2$ and real constant $c_3$.
It is a simple matter to show that these choices of the two sections
are consistent with the supersymmetry variations of SM-I and SM-II
multiplets, respectively.

\sect{V. Yukawa Section Selection Rules}

This brings us straight away to the actual Yukawa-type $n = 1$ terms.
We begin our analysis by a simple enumeration of all such actions.
A convenient notational device for this purpose is provided by
the introduction of a ``model vector'' of the form $({\rm SM},~
{\rm SM}'|~ {\rm MSM}$). The first entry takes on values I,..., IV
labelling which scalar multiplet is used. The second entry takes
on the same values for the same purpose.  The final entry takes
on the same values but indicates which minus spinor multiplet appears.
Following the construction given by Witten, the first two entries
must be chosen to be different.  Finally, the SU(2) symmetry
(that ultimately arises from (4,0) supergravity) places some
restrictions on which minus spinor can appear coupled to particular
pairs of scalar multiplets. When all of this is taken into account,
we find that there are only twelve possibilities to consider.
$$ \eqalign{
({\rm I},~ {\rm II} |~ {\rm I} ) ~~~~~~ & ({\rm III},~ {\rm IV} |~
{\rm I} )  \cr
({\rm I},~ {\rm II} |~ {\rm II} ) ~~~~~~ & ({\rm III},~ {\rm IV} |~
{\rm II} )  \cr
({\rm I},~ {\rm III} |~ {\rm III} ) ~~~~~~ &  ({\rm I},~ {\rm IV} |~
{\rm III} ) ~~~~~~ ({\rm II},~ {\rm III} |~ {\rm III} ) ~~~~~~
({\rm II},~ {\rm IV} |~ {\rm III} ) \cr
({\rm I},~ {\rm III} |~ {\rm IV} )  ~~~~~~ & ({\rm I},~ {\rm IV} |~
{\rm IV} ) ~~~~~~ ({\rm II},~ {\rm III} |~ {\rm IV} ) ~~~~~~
({\rm II},~ {\rm IV} |~ {\rm IV} )
  ~~~ } \eqno(5.1) $$

As long as the sections transform as the scalar multiplet that is
the fermionic dual of the spinor multiplet in each of the actions,
(4,0) supersymmetry will be maintained.  This is a critical point!
The section must not only provide a representation of (4,0)
supersymmetry. It must also be in the fermionic dual representation
of the spinor multiplet!  Thus, for example, ($ r_{\cal A}$, $t_{\cal
B}$, $s_+ {}^i$) must transform like the components of an SM-I type
multiplet. Of course, similar statements must be true about the other
corresponding terms in the other actions.   Since we are only concerned
with the $n = 1$ Yukawa terms, $s_+ {}^i$, $ r_{\cal A}$ and $t_{\cal
B}$ can only depend on monomials of degree (1,1) and consistent with
the first line of (5.1). We have investigated these condition
and we can find no non-trivial solutions! In other words, using
all-known manifest (4,0) formulations implies the impossibility
of writing Yukawa terms!

\sect{VI. (1, 0) and On-shell Supersymmetry Analysis}

Since we have found the surprising and striking result that the use
of {\it {all known}} manifest (4,0) supermultiplets leads to the
impossibility of constructing a model along the lines outlined in
reference \cite{WTN}, it seems as though only allowing a lower
manifest supersymmetry representation or even an on-shell representation
might permit such a construction. For example, we can attempt a (1,0)
superfield formulation. Fortunately, this precise problem has been
studied in great detail previously \cite {Hull}. In fact, the portion
of the work of \cite {WTN} that contains the discussion of including
the minus spinor fields is covered as a special case of the more general
work of \cite {Hull}. Witten's equation (2.10) can be recognized as a
special case of (2.5) (or (3.1)) in the first (or second) work of
reference \cite {Hull}.  Utilizing these previous analyses (after
modifying them to accommodate for two commuting sets of quaternionic
complex structure (see appendix B)), we find that only in the
case of on-shell supersymmetry can a model as described in
reference \cite {WTN} be consistent.

\sect{VII. Summary}

One point we have found is a remarkable and long overlooked fact
in the area of 2D, (4,0) sigma models. Namely, the existence
of variant representations implies that there is a great diversity of
representations for multiplets and the actual construction of
supersymmetric invariant potentials depends crucially on pairing
supersymmetrically dual representations.

In this paper, we have solved a problem that was suggested by
the work in reference \cite{WTN}.  We have seen that the ``missing''
auxiliary fields (absent in reference \cite{WTN}) have greatly
facilitated the analysis of what possible actions may be taken as
the starting point in the most general manifestly (4,0) supersymmetric
action of (4,0) scalar and spinor multiplets.  The most surprising
result of this analysis is that the construction of Witten lies
outside this category of models!

This raises some very interesting questions with regard to the
quantum renormalization behavior of these models.  Within the usual
manifestly (4,0) supersymmetric models, it is possible to derive
non-renormalization theorems based on the the fact that manifestly
(4,0) supersymmetric models are always equivalent to {\underline
{unconstrained}} superfield formulations of these theories. Superfield
perturbation theory lies at the heart of the proofs of the
non-renormalization theorems. Quantum supergraphs require unconstrained
superfield formulations.

On-shell supersymmetric realizations cannot rely on their equivalence
to an unconstrained superfield formulations. In fact, on-shell
supersymmetric realizations cannot be written in terms of
unconstrained superfields. The quantum mechanical behavior of
such theories may be quite different from superfield theories. Thus,
it is of interest to investigate further the quantized versions of
the models of \cite {WTN}.  In fact, this is a very general
question that has not been investigated previously to our
knowledge. Stated most succinctly this question reduces to:
``Is the quantum mechanical behavior of an on-shell
supersymmetric representation always the same as the quantum
mechanical behavior of an off-shell supersymmetric
representation?'' This suggests an avenue requiring future study.

Since we have used a manifest (4,0) supersymmetric formulation, it
is straight forward to couple our matter systems to (4,0) supergravity.
This may prove to be an interesting exercise.  The reason for this
is that although all of our models appear equivalent at the level
of rigid supersymmetry, there are very great differences in the
presence of local (4,0) supersymmetry.  For example, the scalar
and spinors in the SM-I and SM-II multiplets couple in a very
different manner to the (4,0) SU(2) supergravity gauge fields than
do those in the SM-III or SM-IV multiplets.  This also raises the
question of whether the models of \cite {WTN} can be coupled to
supergravity.  There are cases in the literature of on-shell
representations that cannot be coupled to supergravity. Pursuing
this question provides yet another interesting avenue for
future study. \newline $ {~~} $ \newline $ {~~} $
\newline $ {~~} $ \newline $ {~~} $ \newline $ {~~} $
\newline $ {~~} $ \newline $ {~~} $  \newline $ {~~} $

\noindent {\bf{Acknowledgement} }
\indent \newline
S. J. G. wishes to acknowledge E. Witten for turning our attention
to this class of problems.

\newpage

\noindent{{\bf {APPENDIX A: Conventions and Definitions }}}

In heterotic theories, we always adhere to ``helicity index conventions''
that were first established in reference \cite{BMG} and modified in
reference \cite{Gris}.  Thus, a single $+$ ($-$) sign denotes helicity
plus (minus) one-half.  We denote spinors by $\psi_+$ as an example.
In these conventions, the components of a vector must have either
two $+$ or $-$ indices.  Rather than writing two such indices, we
``double them up'' by using the symbols ${}^\dpx$ or ${}^\DM$.  Thus
vectors are typically denoted by $A_{\dpx}$ or $A_{\DM}$. The helicity
index conventions on such vectors are perfectly equivalent to writing
vectors in terms of their light-cone coordinates.  This type of convention
also has the added advantage that by simply counting the number of helicity
indices on a quantity, we can distinguish whether it transforms as a boson
or fermion under the 2D Lorentz group.  Typically bosons have even numbers
of such indices and fermions have odd numbers. Throughout this paper, we
generically use $i,~j,~...$ to denote the components of the defining
representation of SU(2).

\noindent{{\bf {APPENDIX B: Real Formulation of (4,0) Multiplets }}}

It may be useful for future applications to re-write some of our results
in terms of only real fields. It is obvious that at the bottom of all
of the (4,0) multiplets discussed in this paper, there are four real
bosons and four real (Majorana-Weyl) spinors. The supersymmetry
variations of these multiplets can thus be expressed in totally
real form.  For this purpose, we will denote the four scalar fields
in any of the scalar multiplets by $\varphi_A$ with $A = 1,2,3,4$.
Similarly, we introduce four real spinors denoted by ${\Psi}^- {}_{
\hat A}$ with ${\hat A} = 1,2,3,4$. In order to have a (4,0)
supersymmetry, a set of supersymmetry variations can take the form,
$$ \d_Q \varphi_A ~=~ i \a^{+ ~ p} ( {\rm L}_p )_A {}^{\hat A}
{\Psi}^- {}_{\hat A}   ~~~, ~~~ \d_Q {\Psi}^- {}_{\hat A}
 ~=~   \a^{+ ~ p} ( {\rm R}_p )_{\hat A} {}^A \pa_{\dpx} \varphi_A
\eqno(B.1) $$
written in terms of four real constant Grassmann parameters $\a^{+ ~p}$.
In order to form a (4,0) supersymmetry algebra, the real quantities $(
{\rm L}_p )_A {}^{\hat A}$ and $ ( {\rm R}_p )_{\hat A} {}^A$ must satisfy
$$
( {\rm L}_p )_A {}^{\hat A} ( {\rm R}_q )_{\hat A} {}^B ~+~ ( {\rm L}_q )_A
{}^{\hat A}
( {\rm R}_p )_{\hat A} {}^B ~=~ - 2 \d_{p  q} ~ ({\rm I})_A {}^B ~~~, $$
$$
( {\rm R}_p )_{\hat A} {}^A ( {\rm L}_q )_A {}^{\hat B} ~+~ ( {\rm R}_q
)_{\hat A} {}^A
( {\rm L}_p )_A {}^{\hat B}  ~=~ - 2 \d_{p  q} ~ ({\rm I})_{\hat A} {}^{\hat B}
 ~~~.
\eqno(B.2) $$
In other words, the ${\rm L}$-matrices and ${\rm R}$-matrices are generalized
$4 {\rm x} 4$ Pauli matrices.  Thus, to express the SM-I theory and the SM-II
theory in real notation, it is enough to specify the ${\rm L}$-matrices and
${\rm R}$-matrices associated with each multiplet.  A simple calculation
reveals
that there exists a basis in which the SM-I multiplet is associated with the
set
$$ \eqalign{
{\rm L}_1 &=~ i \s^1 \otimes \s^2 ~~~~~~; ~~~~~ {\rm R}_1 ~=~ i \s^1
\otimes \s^2 ~~~ ;  \cr
{\rm L}_2 &=~ i \s^2 \otimes {\rm I} ~~~~~~~~; ~~~~~ {\rm R}_2 ~=~ i \s^2
\otimes
{\rm I} ~~~ ;  \cr
{\rm L}_3 &=~ - i \s^3 \otimes \s^2 ~~~ ; ~~~~~ {\rm R}_3 ~=~ - i \s^3
\otimes \s^2 ~~~ ;  \cr
{\rm L}_4 &=~ - {\rm I} \otimes {\rm I} ~~~~~~~~; ~~~~~ {\rm R}_4 ~=~  {\rm I}
\otimes {\rm I} ~~~~~, } \eqno(B.3) $$
and the SM-II multiplet is associated with
$$ \eqalign{
{\rm L}_1 &=~ i \s^2 \otimes \s^3 ~~~~; ~~~~~ {\rm R}_1 ~=~ i \s^2
\otimes \s^3 ~~~ ;  \cr
{\rm L}_2 &=~ - i {\rm I} \otimes \s^2 ~~~; ~~~~~ {\rm R}_2 ~=~ - i {\rm I}
\otimes \s^2 ~~~ ;  \cr
{\rm L}_3 &=~ i \s^2 \otimes \s^1 ~~~ ; ~~~~~ {\rm R}_3 ~=~ i \s^2
\otimes \s^1 ~~~ ;  \cr
{\rm L}_4 &=~ {\rm I} \otimes {\rm I} ~~~~~~~~; ~~~~~ {\rm R}_4 ~=~ -  {\rm I}
\otimes {\rm I} ~~~~~. } \eqno(B.4) $$
These generalized Pauli matrices resemble complex structures. In fact, there
is a relation between complex structures and these matrices.  If we define
$ (f_p)_A {}^B \equiv ({\rm L}_p {\rm R}_r)_A {}^B$  for fixed $r$ not equal
to $p$, it can be seen that $f_p$ defines a triplet of complex structures.
(The same follows if $ (f_p)_{\hat A} {}^{\hat B} \equiv ({\rm R}_p {\rm
L}_r)_{\hat A} {}^{\hat B}$.)  Equivalently, $(f_{p q})_A {}^B \equiv
\fracm 12 ({\rm L}_p {\rm R}_q - {\rm L}_q {\rm R}_p)_A {}^B$ and $(f_{p
q})_{\hat A} {}^{\hat B} \equiv  \fracm 12 ({\rm R}_p {\rm L}_q - {\rm
R}_q {\rm L}_p)_{\hat A} {}^{\hat B}$ (for unrestricted $p$ and $q$) also
define triplets of complex structures.  Finally, we point out that if we
use $f_p$ to denote the complex structures associated with SM-I and use
${\tilde f}_q$ to denote the complex structures associated with SM-II,
then $[ f_p~,~ {\tilde f}_q ] = 0$ (or eqivalently $ [ f_{p  q}~,~
{\tilde f}_{r s} ] = 0$).

For the SM-III and SM-IV multiplets we will denote the four scalar
fields by $\varphi_{\hat A}$ and the four real spinors by ${\Psi}^-
{}_A$. In order to have (4,0) supersymmetry, the set of supersymmetry
variations take the forms,
$$ \d_Q \varphi_{\hat A} ~=~ i  \a^{+ ~ p} ( {\rm R}_p )_{\hat A} {}^A
{\Psi}^- {}_A   ~~~, ~~~ \d_Q {\Psi}^- {}_A
 ~=~   \a^{+ ~ p} ( {\rm L}_p )_A {}^{\hat A} \pa_{\dpx} \varphi_{\hat A}
{}~~~. \eqno(B.5) $$
There exists a 1D, non-local duality transformation by which we can actually
derive (B.5) starting from (B.1).

Finally very similar results follow for the spinor multiplets. In real
notation MSM-I and MSM-II take the form (below $ {\rm F} {}_{\hat A}$
denote the auxiliary fields),
$$ \d_Q {\Psi}^+{}_A ~=~ i  \a^{+ ~ p} ( {\rm L}_p )_A {}^{\hat A}
{\rm F} {}_{\hat A}   ~~~, ~~~ \d_Q {\rm F} {}_{\hat A}
 ~=~   \a^{+ ~ p} ( {\rm R}_p )_{\hat A} {}^A \pa_{\dpx} {\Psi}^+{}_A ~~~,
\eqno(B.6) $$
with MSM-I and MSM-II associated with (B.3) and (B.4), respectively. For
MSM-III and MSM-IV we have
$$ \d_Q {\Psi}^+{}_{\hat A} ~=~ i  \a^{+ ~ p} ( {\rm R}_p )_{\hat A} {}^A
{\rm F} {}_A   ~~~, ~~~ \d_Q {\rm F} {}_A ~=~   \a^{+ ~ p} ( {\rm L}_p )_A
{}^{\hat A} \pa_{\dpx} {\Psi}^+{}_{\hat A} ~~~,
\eqno(B.7) $$
with MSM-III and MSM-IV associated with (B.3) and (B.4), respectively.

\newpage

\end{document}